# Realizing an Epitaxial Stanene with an Insulating Bandgap

Yunyi Zang, Tian Jiang, Yan Gong, Zhaoyong Guan, Menghan Liao, Zhe Li, Lili Wang, Wei Li, Canli Song, Ding Zhang, Yong Xu*, Ke He*, Xucun Ma, Shou-Cheng Zhang, and Qi-Kun Xue

Dr. Y. Zang, Prof. T. Jiang, Y. Gong, Dr. Z. Guan, M. Liao, Z. Li, Prof. L. Wang, Prof. W. Li, Prof. C. Song, Prof. D. Zhang, Prof. Y. Xu, Prof. K. He, Prof. X. Ma, Prof. Q.-K. Xue

State Key Laboratory of Low Dimensional Quantum Physics, Department of Physics, Tsinghua University, Beijing 100084, P. R. China

Email: yongxu@tsinghua.edu.cn; kehe@tsinghua.edu.cn

Prof. L. Wang, Prof. W. Li, Prof. C. Song, Prof. D. Zhang, Prof. Y. Xu, Prof. K. He, Prof. X. Ma, Prof. Q.-K. Xue

Collaborative Innovation Center of Quantum Matter, Beijing, P. R. China

Prof. T. Jiang

College of Optoelectronic Science and Engineering, National University of Defense Technology, Changsha 410073, P. R. China

Prof. Y. Xu

RIKEN Center for Emergent Matter Science (CEMS), Wako, Saitama 351-0198, Japan

Prof. S.-C. Zhang

Department of Physics, McCullough Building, Stanford University, Stanford, California 94305-4045, USA

The great success of graphene research[1,2] has stimulated tremendous effort to search for other innovative two-dimensional (2D) materials such as silicene,[3] germanene,[4] stanene,[5] phosphorene,[6] borophene,[7] and transition-metal dichalcogenides.[8] The prosperous and diverse 2D materials family constitutes an arsenal for exploring new-concept electronic devices as well as fundamental mysteries of condensed matter.[9] A particularly interesting one of them is stanene—a single atomic layer of gray tin (α-Sn) in a honeycomb lattice structure similar to graphene.[10] Different from graphene, stanene is characterized by its structural buckling and large atomic mass which contribute to strong spin-orbit coupling (SOC), offering an ideal platform to explore topology-related physics and electronics.[11] Stanene and its derivatives have been predicted to be 2D topological insulators (TIs)[12,13] with the bulk bandgap up to several hundred meV. Many other exotic properties, including enhanced thermoelectricity,[14] near-room-temperature quantum anomalous Hall effect,[15] and topological superconductivity,[16] are also expected in stanene-based materials, making them useful for various applications ranging from electronics, spintronics to quantum computation.[11]

In a breakthrough experiment in 2015, single layer stanene was fabricated on $Bi_2Te_3$(111) by molecular beam epitaxy (MBE),[5] which for the first time proved the existence of this theoretically predicted material. Very recently stanene was prepared on Cu(111) and Ag(111) with an unexpectedly flat structure due to strong coupling with the substrates.[17] All these epitaxial stanene samples, however, are grown on metallic substrates ($Bi_2Te_3$ has metallic surface states as a topological insulator) and are driven into metallic phases themselves by interaction with the substrates. Transport studies and electronic applications of stanene are only possible in bulk-insulating stanene samples grown on insulating substrates. In this study, we report the first system satisfying this condition, namely bulk-insulating stanene on Sr-doped PbTe(111). Interestingly, the epitaxial stanene in the present experiment is identified to be a decorated stanene, rather than a sheet of bare stanene. The decorated stanene has passivated $p_z$ orbitals and thus is chemically stable in environment, which does not show Dirac-like linear bands around the K/K' point but displays intriguing low-energy physics around the Γ point.[12] PbTe is a normal insulator with a bandgap of ~0.3 eV and the lattice constant in (111) plane close to that of stanene. We prepared PbTe(111) with MBE by co-evaporating Pb and Te to the surface of $Bi_2Te_3$(111) films on Si(111) (see the schematic of the sample structure in Figure 1a). Figure 1c displays the reflective high energy electron diffraction (RHEED) pattern of a 10 bilayer (BL) PbTe(111) film. The clear and sharp diffraction streaks reveal the high crystalline quality of the film. The in-plane lattice constant is estimated to be 4.52 Å. To fabricate epitaxial stanene, we evaporated Sn onto the PbTe(111) films at 150 K and then annealed the sample up to 400 K. Such low growth and annealing temperatures are necessary to avoid diffusion of Sn atoms into PbTe (see Experimental Section for details). After growth of 1 monolayer (ML) Sn, the RHEED streaks (Figure 1d) remain sharp and exhibit no detectable shift in their positions, which suggests perfect epitaxial growth of Sn on PbTe(111). Figures 1e and 1f display the STM topographies of PbTe(111) before and after growth of 1 ML Sn, respectively. The evolution of the surface morphology clearly shows 2D growth of Sn on PbTe(111) surface (also see Figure S1 of Supporting Information). The depth of the vacancies in the Sn overlayer is 3.6 Å (see the inset of Figure 1f) which is consistent with the thickness of single layer stanene.[5] Furthermore the two-direction RHEED and atomic-resolution STM images illustrate the hexagonal in-plane lattice structure of α-Sn (see Figure S5 of Supporting Information).

Figures 1g-1k display a series of ARPES band maps of Sn/PbTe(111) with Sn coverage increasing from 0 ML to 1 ML around $\bar{\Gamma}$ point of the surface Brillioun zone of PbTe(111) (along cut 1 in Figure 1b). Similar to the early work,[18] the PbTe(111) film is electron-doped and shows a gap of 0.28 eV with the Fermi level cutting the Rashba spin-split surface states that derive from the bulk conduction bands. With increasing Sn coverage, the PbTe bands fade way, and new features appear and finally evolve into two hole bands at $\bar{\Gamma}$ point (Figure 1k).

Figure 2a shows the ARPES spectra of 1 ML Sn/PbTe(111) around $\bar{\Gamma}$ point with larger energy range. One can clearly identify two hole bands, which are designated as H1 and H2, respectively. H1 band is located at 0.32 eV below the Fermi level and 0.34 eV above H2 band. In the spectra around $\bar{\mathrm{K}}$ point (along cut 2 in Figure 1b), a quite blurred band ~0.85 eV below the Fermi level can be distinguished if measurement is taken soon after sample preparation (Figure 2b, the 2nd differential spectra are displayed). After the sample is kept in the MBE chamber (base pressure: ~1×10$^{-10}$ mbar) for over two hours, the band cannot be observed anymore (Figure 2c, the 2nd differential spectra are displayed), meanwhile H1 and H2 bands at $\bar{\Gamma}$ point remain almost unchanged. There is no energy band observed between the Fermi level and H1 band, which suggests that the Sn film is insulating with a bandgap at least 0.32 eV.

To understand the experimental results, we performed first-principles calculations on stanene/PbTe(111) (see details in Experimental Section). As shown in Figure 2d, stanene has a low-buckled honeycomb lattice, constituted by two triangular sublattices stacking together. The bottom sublattice couples with PbTe(111) which was experimentally revealed to be Te-terminated.[18] The upper sublattice is presumed to be decorated by hydrogen atoms. It is reasonable considering that hydrogen (atom or molecule) is ubiquitous and difficult to remove from ultrahigh vacuum MBE chambers.[19] So the chemically active $p_z$ orbitals of bare stanene are fully saturated, which stabilizes the material and leads to sizable band gaps at K/K′, as observed by ARPES (Figure 2c). If the $p_z$ orbitals are not completely passivated (Supporting Information Figure S2), there would be bands observable near the Fermi level at K/K′, which may contribute to the blurred feature observed soon after sample preparation (Figure 2b). In the relaxed geometry (for $a$ = 4.52 Å), the buckling of stanene is 1.0 Å, close to previous data.[11] The vertical distance between the epitaxial stanene and the topmost Te layer is 3.7 Å, consistent with our STM measurement (3.6 Å).

Theoretically, without SOC, the stanene grown on PbTe(111) introduces one electron band and two hole bands around Γ point, which are mainly contributed by Sn-$s$ and Sn-$p_{xy}$ orbitals, respectively (Figure 2e). Noticeably, without SOC the two

$p_{xy}$ hole bands are degenerate at Γ point, and they are split when SOC is turned on (Figure 2f). The two hole bands represent a characteristic feature of stanene[12] which is insensitive to substrate and decoration because of the intactness of in-plane Sn orbitals. They are well consistent with H1 and H2 hole bands observed in ARPES, providing a strong evidence for the existence of single layer stanene. The calculated hole-band splitting at Γ point is 0.49 eV, close to 0.34 eV measured from experiment. According to the calculations, the two characteristic hole bands are further spin-split by Rashba effect (Figure 2f), which was not observed.

As deduced from theory, the energy splitting of the $p_{xy}$ hole bands is determined by SOC and depends weakly on strain effects, which can be tested by experiment. The lattice constant of the epitaxial stanene can be controlled by the thickness of PbTe film. Comparing the RHEED intensity profiles of a 5 BL and a 10 BL PbTe films, we can observe a shift in the diffraction peaks (Figures 3a and 3b). The estimated surface lattice constants of the 5 BL and 10 BL PbTe films are 4.46 Å and 4.52 Å, respectively.[20,21] Stanene always keeps perfect epitaxy with PbTe regardless of its lattice constant. Figures 3c and 3e show the ARPES bandmaps of the single layer stanene films grown on 5 BL and 10 BL PbTe films, respectively. The band structures of the two samples show no obvious difference despite their different lattice constants. The energy splittings between H1 and H2 bands in the two samples are almost the same (Figures 3d and 3f), in excellent agreement with theoretical prediction (the calculated splitting changes by less than 0.01 eV). This observation implies that the valence bands of the epitaxial stanene mainly have $p_{xy}$ component, without an *s-p* band inversion. It implies that the sample might be a topologically trivial insulator.

Checking the band alignment between the epitaxial stanene and PbTe(111) (see the schematic in Figure 4a and the analysis on Figure S3 in Supporting Information), we found that the conduction bands of PbTe cover a large portion of the stanene gap. The Fermi level, although lying in the stanene gap, cuts the substrate conduction bands, which is unfavorable for transport studies and electronic applications of stanene. Doping a small amount of Sr in PbTe can significantly increase its bulk gap up to ~0.7 eV.[22] We prepared Sr-doped PbTe(111) by co-evaporating Sr in MBE growth of PbTe. In the ARPES bandmap of a $Pb_{0.9}Sr_{0.1}Te(111)$ film (Figure 4c), the valence band maximum is 0.65 eV below the Fermi level, and no conduction bands are observed, which suggests a truly insulating substrate resulting from bandgap enhancement. The band structure of stanene grown on $Pb_{0.9}Sr_{0.1}Te(111)$ is almost the same as that of stanene/PbTe(111). The bulk gap of the epitaxial stanene is now located in the substrate gap (see the schematic in Figure 4b). The stanene/$Pb_{0.9}Sr_{0.1}Te(111)$ structure is thus expected to show native properties of stanene in transport measurements.

Although the obtained epitaxial stanene is likely to be a topologically trivial insulator, one can drive it into the 2D TI phase by increasing the film thickness or the lattice constant.[12,23] From the ARPES data of a bilayer and a triple-layer stanene films (Figure S4 and Ref. 25), we observed additional hole bands crossing the Fermi level, which could lead to an *s-p* band inversion at higher energy. Further studies are needed to reveal their topological properties. Since the lattice constant of epitaxial stanene can be tuned by the PbTe substrate, by using similar substrate materials with larger lattice constants, such as EuTe, SrTe, BaTe, or their alloys, one may further expand the lattice of epitaxial stanene, driving it into the 2D TI phase.[24]

In summary, we have obtained bulk insulating epitaxial stanene by using Sr-doped PbTe(111) as substrate. Our studies revealed that the $p_z$ orbitals of stanene are automatically passivated even in an ultrahigh vacuum chamber, rendering the properties of decorated stanene insensitive to environment. The realization of truly insulating stanene samples lays a foundation for future experimental studies of various quantum phenomena and electronic applications of stanene.

**Experimental Section**

The experiments were performed in an ultrahigh vacuum system (base pressure:

$1\times10^{-10}$ mbar) consisting of MBE (Omicron), STM (Omicron) and ARPES (VG-Scienta) facilities. All the elements were evaporated from standard Knudsen cells. The growth of stanene on PbTe(111) was carried out in a two-step procedure. First, Sn atoms were deposited onto PbTe(111) with the substrate at ~150 K. And the film was then annealed from 150 K to ~400 K, which improves the film quality. Higher annealing temperature results in significant change in the surface morphology and the band structure, probably due to inter-diffusion between Sn and PbTe. The flux of Sn is calibrated by STM (Supporting Information Figure S1). STM measurements were carried out at 77K. ARPES measurements were carried out at 90 K with a Scienta R4000 spectrometer and a Gammadata helium discharge lamp (He-Iα, $h\nu =$ 21.2 eV).

Computation Methods: First-principles density functional theory calculations were performed by the Vienna *ab initio* simulation package, using the projector-augmented-wave potential, Perdew-Burke-Ernzerhof (PBE) exchange-correlation functional and planewave basis with an energy cutoff of 400 eV. Stanene/PbTe(111) was modeled by the periodic slab approach using two Pb-Te bilayers with the bottom bilayer fixed and saturated by fluorine for removing the dangling bonds, a vacuum layer of over 12 Å and an 12×12×1 Monkhorst-Pack *k* grid. The ultrathin PbTe slab was used to simulate the local chemical bonding of stanene/PbTe(111), which gives a band gap of PbTe larger than reality, facilitating the analysis of stanene bands. Stanene/PbTe(111) was saturated by hydrogen for fully passivating the $p_z$ orbitals, but the calculated band structure is insensitive to the saturation group. In contrast, ultraflat stanene on Cu(111) and Ag(111) interacts strongly with the substrate (adsorption energy > 1.0 eV/atom)[17], which significantly passivates the $p_z$ orbitals and thus makes the material insensitive to the residual gas in experiment. The strain effect applied by the substrate was studied by using different surface lattice constants determined experimentally, *a* = 4.46 and 4.52 Å for 5BL and 10BL PbTe(111) thin films, respectively. The spin-orbit coupling was included in the self-consistent calculations of electronic structure.

**Supporting Information**

Supporting Information is available from the Wiley Online Library or from the author.

**Acknowledgements**

This work is supported by the National Natural Science Foundation of China (grant No. 51661135024) and Ministry of Science and Technology of China (grant No. 2017YFA0303303). Y.X. acknowledges support from Tsinghua University Initiative Scientific Research Program and the National Thousand-Young-Talents Program. S.-C. Z. is supported by the U.S. Department of Energy, Office of Basic Energy Sciences, Division of Materials Sciences and Engineering under Contract No. DE-AC02-76SF00515.

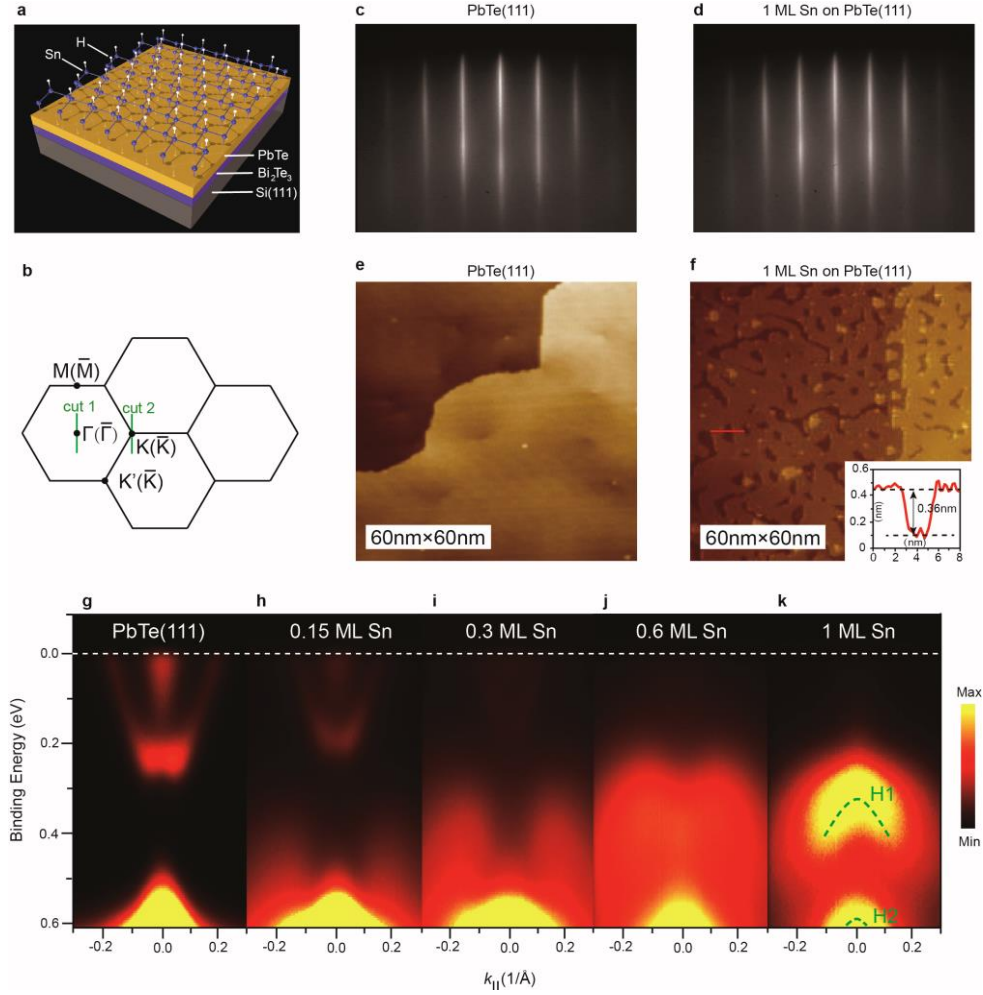

**Figure 1. MBE growth of stanene on PbTe(111). a**) Schematic of the sample structure. The $p_z$ orbitals of stanene are passivated, possibly by hydrogen. **b**) Schematic of the Brillioun zones of stanene (with the high-symmetry points Γ, K/K' and M) which exactly coincide with the surface Brillioun zones of PbTe(111) (with the high-symmetry points $\bar{\Gamma}$, $\bar{K}$, $\bar{M}$) due to perfect epitaxy between stanene and PbTe(111). **c,d**) RHEED patterns of a 10 BL PbTe(111) film (**c**) and 1 ML Sn grown on a 10 BL PbTe(111) film. **e,f**) STM images (both 60 nm × 60 nm) of a 10 BL PbTe(111) film (**e**) and 1 ML Sn grown on 10 BL PbTe(111) film (**f**). The inset of **f** shows the line profile along the red line. **g-k**) ARPES bandmaps of Sn grown on PbTe(111) with coverage of 0 ML (**g**), 0.15 ML (**h**), 0.3 ML (**i**), 0.6 ML (**j**), and 1 ML (**k**) around $\bar{\Gamma}$ point, respectively.

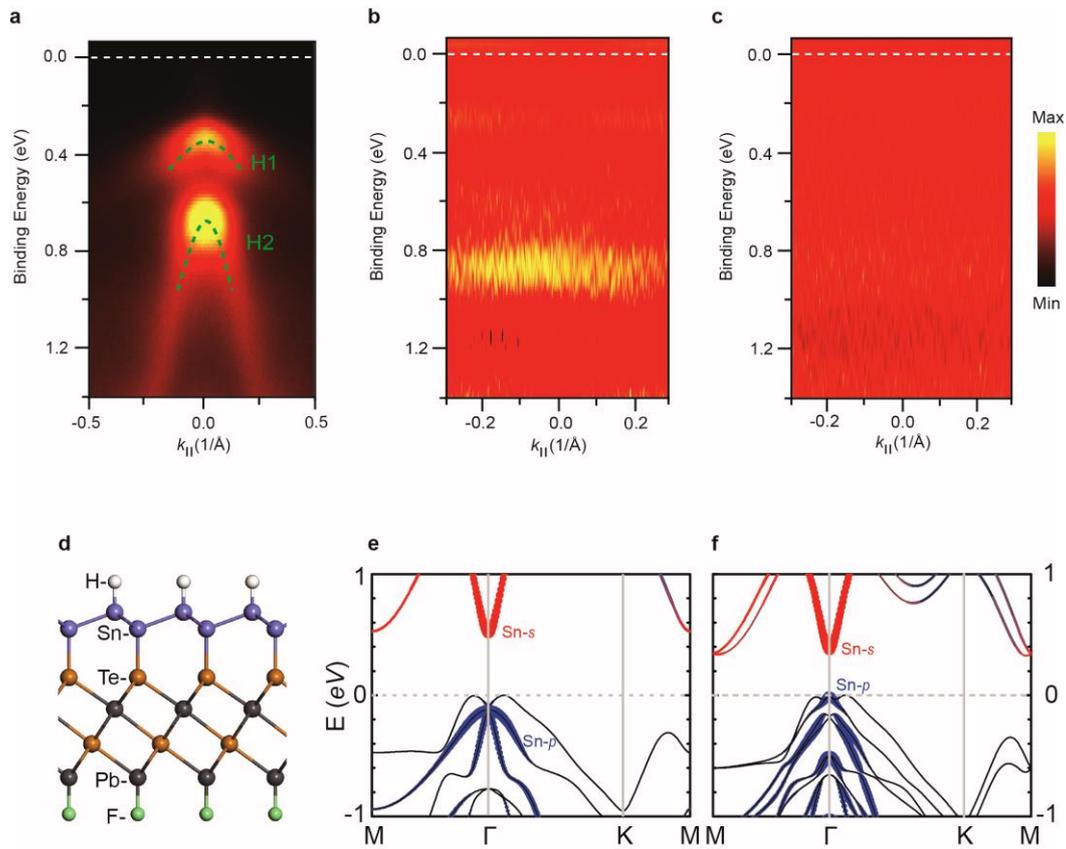

**Figure 2. Energy band structure of stanene grown on PbTe(111). a**) ARPES bandmap of single layer stanene epitaxied on PbTe(111) around $\bar{\Gamma}$ point along cut 1 in Figure 1b. **b**) ARPES bandmap (2nd differential data) of as-prepared stanene epitaxied on PbTe(111) around $\bar{K}$ point along cut 2 in Figure 1b. **c**) ARPES bandmap (2nd differential data) of stanene epitaxied on PbTe(111) two hours after preparation around $\bar{K}$ point along cut 2 in (**b**). **d**) Atomic model of stanene/PbTe(111). **e,f**) Calculated band structures of stanene/PbTe(111) ($a$ = 4.52 Å) excluding (**e**) and including (**f**) the spin-orbit coupling. Red (blue) color highlights the contributions from Sn-$s$ (Sn-$p$) orbital, and energy is referenced to the valence band maximum.

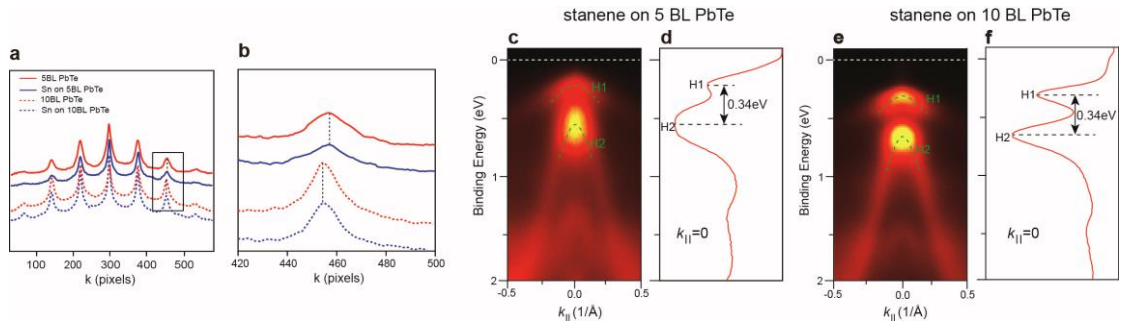

**Figure 3. Band structures of the epitaxial stanenes of different lattice constants. a**) The intensity profiles of the RHEED patterns of a 5 BL and a 10 BL PbTe(111) films with and without stanene grown on it. **b**) A close view of the intensity profiles shown in the black box in (**a**). **c,e**) ARPES bandmaps of stanenes grown on 5 BL (**c**) and 10 BL (**e**) PbTe(111) films. **d,f**) Normal photoemission spectra ($k_{//}=0$) of stanenes grown on 5BL (**d**) and 10BL (**f**) PbTe(111) films.

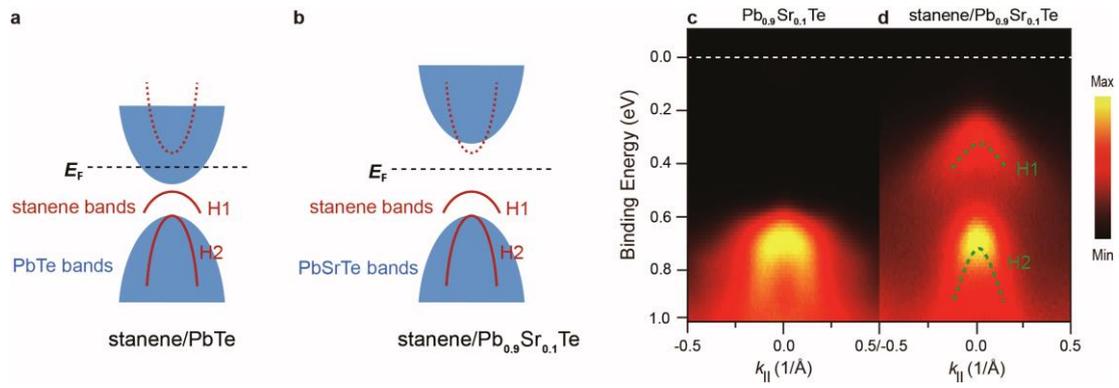

**Figure 4. Band structure of stanene grown on Sr-doped PbTe(111)**. **a**) Schematic of the band structure of stanene grown on PbTe(111). **b**) Schematic of the band structure of stanene grown on $Pb_{0.9}Sr_{0.1}Te(111)$. **c**) ARPES bandmap of 10 BL $Pb_{0.9}Sr_{0.1}Te(111)$. **d**) ARPES bandmap of single layer stanene grown on 10 BL $Pb_{0.9}Sr_{0.1}Te(111)$.

# Supporting Information

Figures S1a, S1b and S1c show the STM images of 0 ML, 0.3 ML and 1 ML Sn grown on a 10 BL PbTe(111) film, respectively. The STM images show the typical two-dimensional growth mode. The flux of Sn source was calibrated by the Sn coverages shown in the STM images.

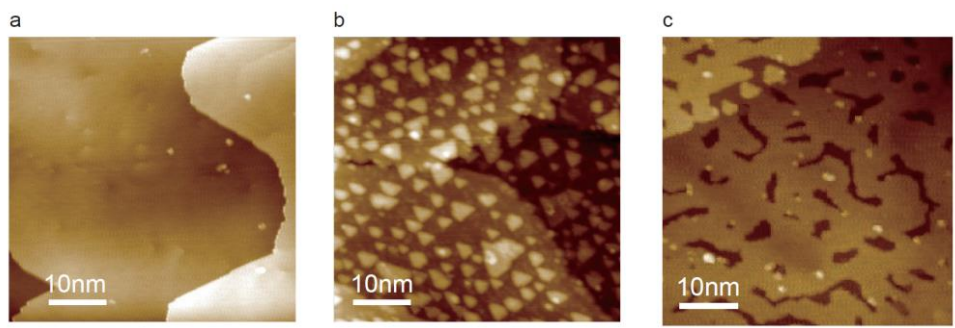

**Figure S1**. STM images of 0 ML (**a**), 0.3 ML (**b**) and 1 ML (**c**) Sn grown on PbTe(111).

Figure S2 shows the calculated band structures of single layer stanene epitaxied on PbTe(111) without being passivated by hydrogen atoms. The non-passivated $p_z$ bonds make the stanene metallic and contribute to the bands at K(K') points.

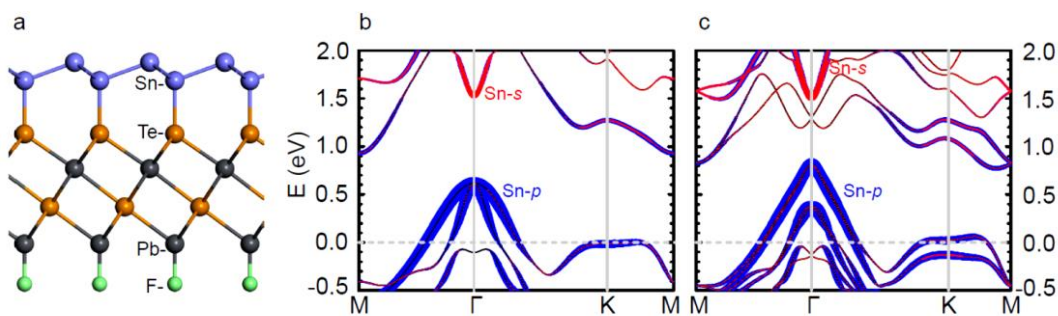

**Figure S2. a**) Atomic structure model of stanene epitaxied on PbTe(111) ($a = 4.52$ Å) without being passivated by hydrogen atoms. **b,c**) The calculated band structure without (**b**) and with (**c**) the SOC of non-passivated stanene/PbTe(111). Red (blue) color highlights the contributions from Sn-$s$ (Sn-$p_{xy}$) orbital.

Figures S3a and S3b display the bandmaps of PbTe(111) and the epitaxial stanene on PbTe(111) in the smaller energy range. Figures S3c and S3d display the bandmaps of PbTe(111) and the epitaxial stanene on PbTe(111) in a larger energy range. In both Figures S3c and S3d, we identify a hole band which becomes blurred in Figure S3d (indicated by the blue dashed line). Obviously, it is a PbTe valence band. The position of the band barely changes with the stanene grown on PbTe, which suggests negligible charge transfer between stanene and PbTe. Therefore, there is little energy offset between the PbTe bands shown in Figure S3a and the stanene bands shown in Figure S3b. We can see that the minimum of the Rashba surface states of PbTe is only ~100 meV above the maximum of H1 band of stanene. The conduction band minimum of PbTe has been shown to be very close to the Rashba surface states (Ref. 18). So the epitaxial stanene/PbTe(111) as a whole only has a bulk gap of ~100 meV and is heavily electron-doped.

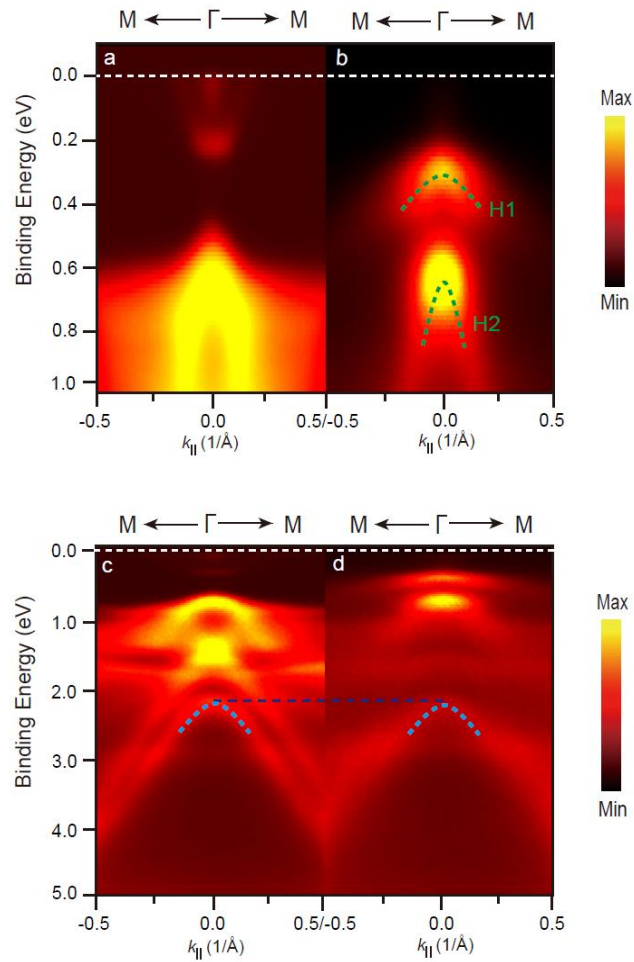

**Figure S3**. **Energy band alignment between epitaxial stanene and PbTe(111). a,c)** ARPES bandmaps of 10 BL PbTe(111) with smaller (**a**) and larger (**c**) energy ranges. **b,d**) ARPES bandmaps of stanene epitaxied on 10 BL PbTe(111) smaller (**b**) and larger (**d**) energy ranges.

Figures S4a and S4b display the ARPES bandmaps of a bilayer and a triple-layer stanenes grown on PbTe(111), respectively. Compared with the band structure of single layer stanene (Figure 2a), additional hole bands (indicated by the red dash line) are observed crossing the Fermi level.

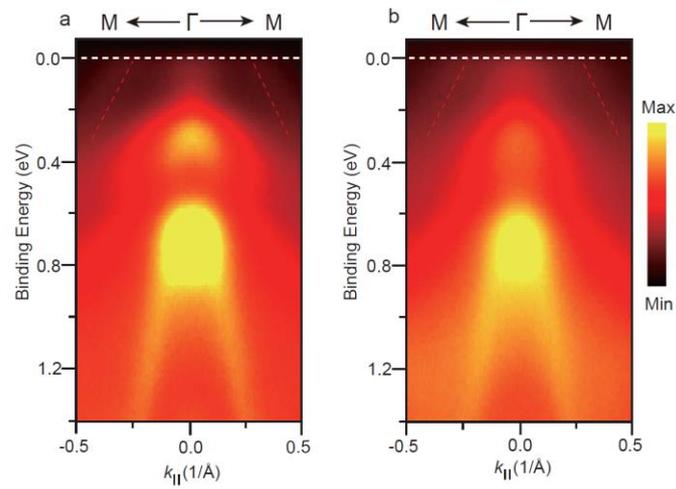

**Figure S4**. ARPES bandmaps of a bilayer (**a**) and a triple-layer (**b**) stanenes on 10 BL PbTe(111).

Figure S5a and S5b show the RHEED patterns of α-Sn film grown on PbTe(111) substrate along the $[11\bar{2}]$ and $[1\bar{1}0]$ directions, respectively. The separations of the RHEED stripes from the two perpendicular directions show a ratio of $\sqrt{3}:1$, demonstrating the six-fold symmetry of epitaxial α-Sn film. Furthermore, the atomic STM image of α-Sn film (Figure S5c) and the Fourier transformed image (The Inset of Figure S5c) illustrate the hexagonal in-plane lattice structure of the α-Sn surface.

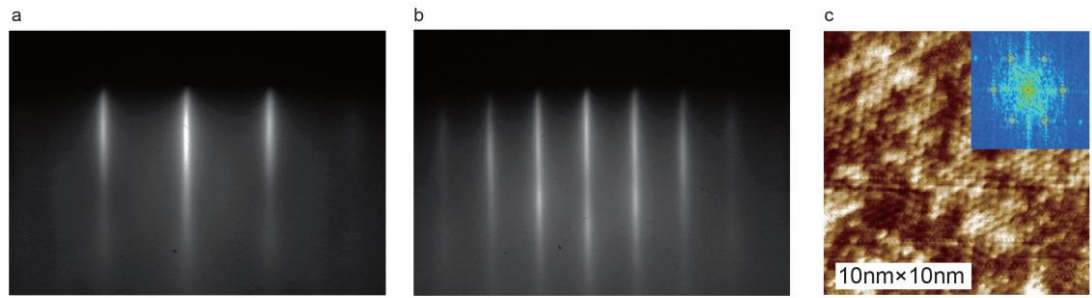

**Figure S5**. RHEED patterns of monolayer α-Sn film along **a)** $[11\bar{2}]$ and **b)** $[1\bar{1}0]$ directions. **c)** Atomic-resolution STM of monolayer α-Sn film. The inset shows the Fourier transformed image.